# Persistent spin texture enforced by symmetry

L. L. Tao and Evgeny Y. Tsymbal

*Department of Physics and Astronomy & Nebraska Center for Materials and Nanoscience,*
*University of Nebraska, Lincoln, Nebraska 68588, USA*

**Persistent spin texture (PST) is the property of some materials to maintain a uniform spin configuration in the momentum space. This property has been predicted to support an extraordinarily long spin lifetime of carriers promising for spintronics applications. Here we predict that there exist a class of non-centrosymmetric bulk materials where the PST is enforced by the non-symmorphic space group symmetry of the crystal. Around certain high symmetry points in the Brillouin zone, the sublattice degrees of freedom impose a constraint on the effective spin-orbit field, which orientation remains independent of the momentum and thus maintains the PST. We illustrate this behavior using density-functional theory calculations for a handful of promising candidates accessible experimentally. Among them is the ferroelectric oxide $BiInO_3$ – a wide band gap semiconductor which sustains a PST around the conduction band minimum. Our results broaden the range of materials, which can be employed in spintronics.**

In recent years, there has been increasing interest in materials and structures where quantum effects are responsible for novel physical properties, revealing the important roles of symmetry, topology, and dimensionality.[1] Among such quantum materials are graphene, topological insulators, Weyl semimetals, and superconductors. In many cases, the quantum materials derive their properties from the interplay between the electron, spin, lattice, and orbital degrees of freedom, resulting in complex physical phenomena and emergent functionalities.[2] These new functionalities are interesting due to their potential for a continuously evolving field of spintronics.[3]

Regarding the new phenomena, often a special role is played by the spin-orbit coupling (SOC), which on its own has inspired a vast number of predictions, discoveries, and novel concepts.[4] In a system, lacking an inversion center, the SOC results in an effective momentum-dependent magnetic field acting on spin $\sigma$. This field $\Omega(\mathbf{k})$ is odd in the electron's wave vector ($\mathbf{k}$), as was first demonstrated by Dresselhaus[5] and Rashba,[6] so that the effective SOC Hamiltonian can be written as

$$H_{SO} = \Omega(\mathbf{k}) \cdot \sigma, \qquad (1)$$

preserving the time-reversal symmetry. The specific form of $\Omega(\mathbf{k})$ depends on the space symmetry of the system. For example, in case of the $C_{2v}$ point group, the Dresselhaus and Rashba SOC fields can be written as $\Omega_D(\mathbf{k}) = \lambda_D(k_y, k_x, 0)$ and $\Omega_R(\mathbf{k}) = \lambda_R(-k_y, k_x, 0)$, respectively. Such SOC leads to a chiral spin texture of the electronic bands in the momentum space, as shown in Figs. 1a and 1b. The chiral spin textures driven by the SOC can be exploited to create non-equilibrium spin polarization,[7] produce the spin Hall effect,[8] and design a spin field-effect transistor (FET).[9] Recently, these and other related phenomena have received significant attention and led to the emergence of a new field of research – spin-orbitronics.[4]

Although large SOC is beneficial for realizing these phenomena, it plays a detrimental role for the spin life time. In a diffusive transport regime, impurities and defects scatter electrons, changing their momentum and randomizing the spin, due to the momentum-dependent spin-orbit field $\Omega(\mathbf{k})$. This process known as the Dyakonov-Perel spin relaxation[10] reduces the spin life time and thus limits the performance of potential spintronic devices, e.g., the spin FET. A possible way to circumvent this effect is to engineer a structure where the spin-orbit field orientation is momentum-independent.[11] This can be achieved, in particular, if the magnitudes of $\lambda_R$ and $\lambda_D$ are equal, i.e. $\lambda/2 = \lambda_D = \pm\lambda_R$, resulting in a unidirectional spin-orbit field, $\Omega_{PST} = \lambda(k_y, 0, 0)$ or $\Omega_{PST} = \lambda(0, k_x, 0)$, and thus a momentum-independent spin configuration, known as the persistent spin texture (PST) (Fig. 1c).[12]

Under these conditions, electron motion is accompanied by spin precession around the unidirectional spin-orbit field, resulting in a spatially periodic mode known as a persistent spin helix (PSH).[13] The PSH state arises due to the SU(2) spin rotation symmetry, which is robust against spin-independent disorder and renders an ultimately infinite spin lifetime.[14] The PSH has been experimentally demonstrated in the two-dimensional electron gas semiconductor quantum-well structures, such as GaAs/AlGaAs[15,16] and InGaAs/InAlAs,[17,18] where the required condition of equal Rashba ($\lambda_R$) and Dresselhaus ($\lambda_D$) parameters was realized through tuning the quantum-well width, doping level, and applied external electric field.

Despite these advances, a number of difficulties impede the practical application and further experimental studies of these semiconductor heterostructures. Satisfying the stringent condition of equal $\lambda_R$ and $\lambda_D$ parameters is technically non-trivial because it requires a precise control of the quantum-well width and the doping level. Furthermore, due to the small values of these parameters (a few meV Å), efficient spin manipulation by an applied electric field is questionable. Recently, based on first-principles calculations a PST was predicted for a wurtzite ZnO ($10\bar{1}0$) surface[19] and a tensile-strained LaAlO$_3$/SrTiO$_3$ (001) interface.[20] However, for the latter, too large tensile strain (>5%) is required to achieve the desired property, whereas for the former, the SOC energy splitting is too small (~1 meV). It would be desirable to find bulk materials where the PST is a robust intrinsic bulk property. Recently, SnTe (001) thin films have been proposed to realize a PSH.[21]

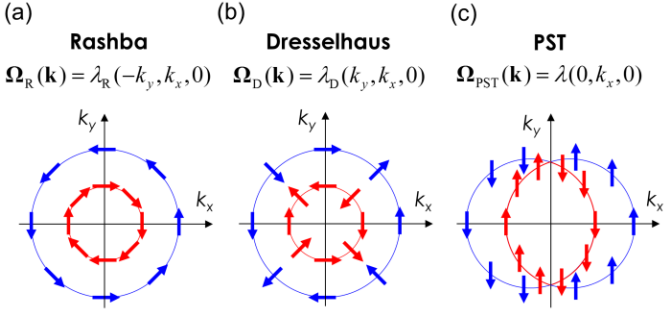

**Fig. 1 Spin texture. (a,b,c)** Spin structure resulting from spin-orbit coupling in a system lacking an inversion center: Rashba (a), Dresselhaus (b), and persistent spin texture (c) configurations. Blue and red arrows indicate spin orientation for the two electronic sub-bands resulting from SOC. Expressions for the respective SOC fields $\mathbf{\Omega}(\mathbf{k})$ are shown. Note that $\mathbf{\Omega}_D$ is represented in the coordinate system with the $x$- and $y$-axes being perpendicular to the mirror planes of an orthorhombic system ($M_x$ and $M_y$ in Fig. 2a).

Here we propose a conceptually different approach to achieve the PST. We demonstrate that there exist a class of non-centrosymmetric bulk materials where the PST is enforced by non-symmorphic space group symmetry of the crystal, i.e. the space group combining point-group symmetry operations with non-primitive translations.[22] Around certain high symmetry points in the Brillouin zone, the sublattice degrees of freedom impose a constraint on the effective spin-orbit field, which orientation remains independent of the momentum and thus maintains PST. The symmetry-enforced PST survives over the large part of the Brillouin zone including band edges, as we demonstrate using density-functional theory (DFT) calculations for BiInO$_3$ and other materials with appropriate crystal group symmetry.

## Results

**Symmetry analysis.** We consider orthorhombic non-symmorphic crystals with broken space inversion symmetry (space groups listed in Table 1).[22] Fig. 2a shows an orthorhombic crystal lattice, which contains the following symmetry operations: (1) the identity operation $E$; (2) glide reflection $\bar{M}_x$ which consists of mirror reflection about the $x = 0$ plane $M_x$ followed by the $(\mu_1 \nu_1 \eta_1)$ translation:

$$\bar{M}_x : (x, y, z) \to (-x + \mu_1, y + \nu_1, z + \eta_1); \quad (2)$$

(3) glide reflection $\bar{M}_y$ which consists of mirror reflection about the $y = 0$ plane $M_y$ followed by the $(\mu_2 \nu_2 \eta_2)$ translation:

$$\bar{M}_y : (x, y, z) \to (x + \mu_2, -y + \nu_2, z + \eta_2); \quad (3)$$

(4) two-fold screw rotation $\bar{C}_{2z}$ which consists of two-fold rotation around the $z$ axis $C_{2z}$ followed by the $(\mu_3 \nu_3 \eta_3)$ translation:

$$\bar{C}_{2z} : (x, y, z) \to (-x + \mu_3, -y + \nu_3, z + \eta_3). \quad (4)$$

Here and below, the translation and reciprocal vectors are given in units of lattice constants and $\mu_i, \nu_i, \eta_i = 0, \frac{1}{2}$ ($i = 1, 2, 3$). The glide (screw) symmetry is reduced to mirror (rotation) symmetry if $\mu_i = \nu_i = \lambda_i = 0$. In addition, we assume that the system exhibits time-reversal symmetry $T$.

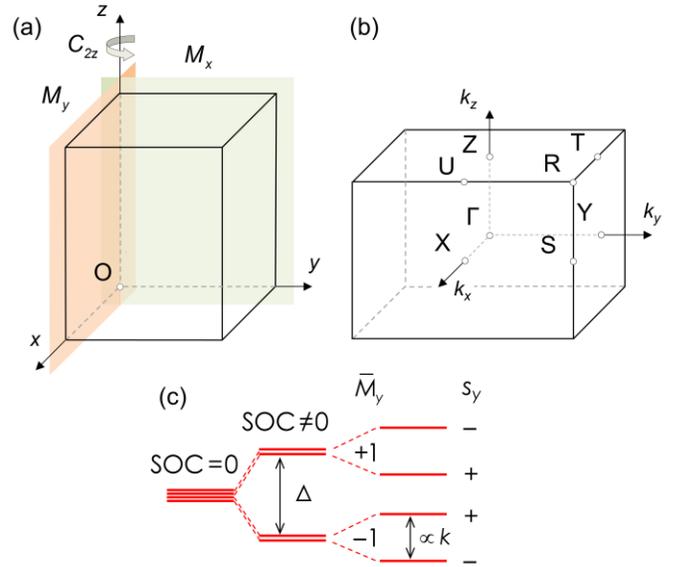

**Fig. 2 Crystal lattice and energy band splitting. (a)** Orthorhombic crystal lattice with symmetry operations indicated. $C_{2z}$ denotes a two-fold rotation operator, and $M_x$ and $M_y$ represent two mirror reflection operators. **(b)** The first Brillouin zone with the high symmetry **k** points indicated: Γ (0, 0, 0), X (π, 0, 0), S (π, π, 0), Y (0, π, 0), Z (0, 0, π), U (π, 0, π), R (π, π, π), and T (0, π, π), the **k** point coordinates are given in units of the reciprocal lattice constants. **(c)** Schematic splitting of the energy levels around the X point. SOC splits the state into two doublets with eigenvalues of $\bar{M}_y = \pm 1$, which are further split into singlets with sign-reversed expectation values of $s_y$. The energy level order labeled by $\bar{M}_y$ and $s_y$ is material dependent.

Now we demonstrate the formation of the PST around the X point $\mathbf{k}=(\pi,0,0)$ in the Brillouin zone of the crystal (Fig. 2b). First, we consider the X-S high symmetry line $\mathbf{k}=(\pi,k_y,0)$. Along this line, the little group of wave vector $\mathbf{k}$ includes the symmetry operators $\bar{M}_x$ and $\Theta \equiv T\bar{M}_y$, as follows from $k_y$ being invariant under the transformations determined by these symmetry operators. Since $T^2=-1$ for a spin-half system, we find $\Theta^2=T^2\bar{M}_y^2=e^{-2i\mu_2\pi}$. Therefore, for the space groups with $\mu_2=\tfrac{1}{2}$, along the X-S line, $\Theta^2=-1$ so that all bands are double degenerate. The doublet states $(\psi_\mathbf{k},\Theta\psi_\mathbf{k})$ form a Kramers's pair.

At the X point, $\bar{M}_y$ commutes with the Hamiltonian of the crystal, i.e. $\left[\bar{M}_y,H\right]=0$, and the doublet $(\psi_X,\Theta\psi_X)$ can be labeled using the eigenvalues of $\bar{M}_y$. Since $\bar{M}_y^2=1$ at this point, we have $\bar{M}_y\psi_X^\pm=\pm\psi_X^\pm$ and $\bar{M}_y\Theta\psi_X^\pm=\pm\Theta\psi_X^\pm$. Thus, by symmetry, there are two conjugated doublets at the X point, $(\psi_X^+,\Theta\psi_X^+)$ or $(\psi_X^-,\Theta\psi_X^-)$, which are distinguished by the $\bar{M}_y$ eigenvalues. Within each of the two doublets, matrix elements of the spin operators $\sigma_x$ and $\sigma_z$ are equal to zero. This is due to the fact that in the spin space $\bar{M}_y$ anti-commutes with $\sigma_x$ and $\sigma_z$, i.e. $\{\bar{M}_y,\sigma_{x,z}\}=0$, which results in $\langle\psi_X^+|\sigma_{x,z}|\psi_X^+\rangle=\langle\psi_X^+|\bar{M}_y^{-1}\sigma_{x,z}\bar{M}_y|\psi_X^+\rangle=-\langle\psi_X^+|\sigma_{x,z}|\psi_X^+\rangle$, and hence $\langle\psi_X^+|\sigma_{x,z}|\psi_X^+\rangle=0$. The similar analysis leads to $\langle\Theta\psi_X^+|\sigma_{x,z}|\Theta\psi_X^+\rangle=0$ and $\langle\psi_X^+|\sigma_{x,z}|\Theta\psi_X^+\rangle=0$. The same conclusion holds for the other doublet $(\psi_X^-,\Theta\psi_X^-)$.

**Table 1.** Classification of orthorhombic space groups with no inversion symmetry according to translation vectors characterized by indices $(\mu_2\nu_1)$. Non-zero spin components in high symmetry points and band degeneracy along high symmetry lines are shown.

| $(\mu_2\nu_1)$ | X | Y | Band degeneracy | Space group No. |
|---|---|---|---|---|
| $(\tfrac{1}{2}0)$ | $s_y$ | - | X-S | 28, 29, 31, 40, 46 |
| $(0\tfrac{1}{2})$ | - | $s_x$ | Y-S | 30, 39 |
| $(\tfrac{1}{2}\tfrac{1}{2})$ | $s_y$ | $s_x$ | X-S and Y-S | 32, 33, 34, 41, 45 |

We see therefore that any state, which represents a linear combination of the states comprising either doublet, i.e. $\psi_\mathbf{k}^\pm=a_\mathbf{k}\psi_X^\pm+b_\mathbf{k}\Theta\psi_X^\pm$ (where $a_\mathbf{k}$ and $b_\mathbf{k}$ are some coefficients), has zero expectation values of $\sigma_{x,z}$ and zero spin components $s_{x,z}=\tfrac{1}{2}\langle\psi_\mathbf{k}^\pm|\sigma_{x,z}|\psi_\mathbf{k}^\pm\rangle=0$. The only non-zero component of the spin is therefore $s_y$. Thus, as long as the two doublets are not mixed, the spin orientation is forced to be along the $y$ direction.

This explains the PST around the X point. At the X point the SOC splits the four-fold degenerate state into two doublets with splitting $\Delta$ and eigenvalues of $\bar{M}_y=\pm 1$, as shown in Fig. 2c. When moving away from this point the perturbation breaks the X point symmetry and further splits the doublets, each into two singlets (unless going along the X-S symmetry line). These states preserve the unidirectional spin texture along the $y$ direction unless the perturbation is so strong that it mixes the doublets. However, due to the perturbation being linear with respect to $\mathbf{k}$ (measured from the X point), there is always a range of $\mathbf{k}$ vectors where it is small compared to the splitting between the doublets. In practice, this range of $\mathbf{k}$ values may be substantial and can span a large portion of the Brillouin zone including the band edges responsible to transport and optical properties in semiconductor materials.

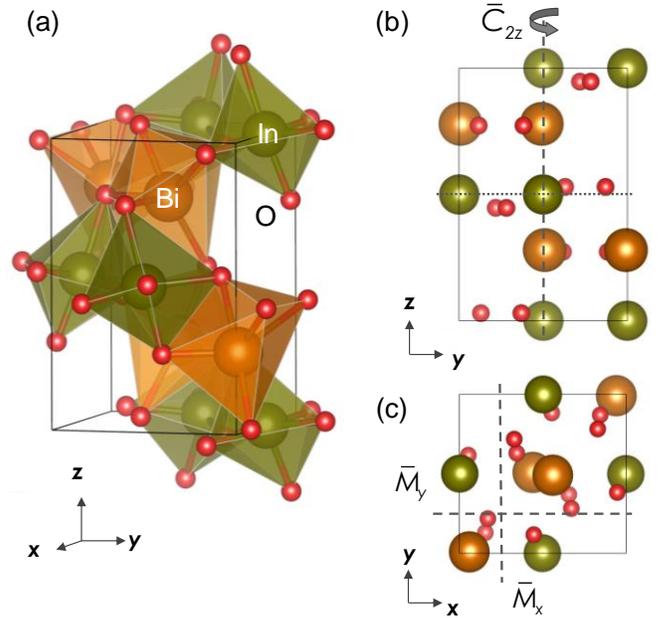

**Fig. 3 Crystal structure of bulk BiInO$_3$.** (**a**) 3D view of the unit cell structure. (**b, c**) View of the crystal structure in the (100) plane (b) and the (001) plane (c). The two-fold screw rotation axis ($\bar{C}_{2z}$) and the glide reflection planes ($\bar{M}_x$ and $\bar{M}_y$) are indicated by the dashed lines. The dotted line indicates a *Pmna* symmetry mirror plane.

A similar analysis applies to the Y point, where $\mathbf{k}=(0,\pi,0)$ (Fig. 2b). The bands are double degenerate along the high symmetry Y-S line, where $\mathbf{k}=(k_x,\pi,0)$, provided that $\nu_1=\tfrac{1}{2}$. The wave functions at the Y point are the eigenstates of the spin component $\sigma_x$. A portion of the Brillouin zone around the Y point maintains the PST with the spin pointing along the $x$ direction. In Table 1 we classify space groups of the orthorhombic crystal system according to the $(\mu_2, \nu_1)$ value and show those spin components $\mathbf{s} = (s_x, s_y, s_z)$ which remain non-zero around the respective high-symmetry points.

**DFT analysis of bulk BiInO$_3$.** In the following, we reinforce our symmetry-based conclusions by performing DFT calculations for a number of bulk compounds, which belong to selected space groups listed in Table 1. Details of the DFT calculations are described in Methods. First, we focus on perovskite BiInO$_3$ (space group No. 33), which has been synthesized experimentally and is stable at ambient conditions.[23] The BiInO$_3$ crystal structure (Fig. 3) belongs to the $Pna2_1$ orthorhombic phase (space group No. 33). The symmetry operations of this group involve the glide reflection $\bar{M}_x$ (2) with $(\mu_1 = \tfrac{1}{2}, \nu_1 = \tfrac{1}{2}, \eta_1 = \tfrac{1}{2})$, the glide reflection $\bar{M}_y$ (3) with $(\mu_2 = \tfrac{1}{2}, \nu_2 = \tfrac{1}{2}, \eta_2 = 0)$ and the two-fold screw rotation $\bar{C}_{2z}$ (4) with $(\mu_3 = 0, \nu_3 = 0, \eta_3 = \tfrac{1}{2})$. The BiInO$_3$ crystal structure is derived from the centrosymmetric GdFeO$_3$-type perovskite structure (*Pnma* group) through polar displacements, which break space inversion symmetry. As seen from Fig. 3a, each bismuth or indium atom in the BiInO$_3$ structure is surrounded by a distorted oxygen octahedron typical for the GdFeO$_3$-type perovskite structure. In addition, there are polar displacements seen, e.g., in Fig. 3b from displacement of Bi$^{3+}$ ions (~0.25 Å) from their symmetric positions with respect to the mirror *Pnma* plane (dotted line in Fig. 3b). The polar displacements yield a finite polarization pointing in the [001] direction. There are two topologically equivalent variants of the space group $Pna2_1$ with opposite polarization (pointing in the [001] or [00$\bar{1}$] directions) indicative to the ferroelectric nature of BiInO$_3$. The calculated polarization is about 33.6 μC/cm$^2$.

Fig. 4a shows the calculated band structure of BiInO$_3$ without SOC along high-symmetry lines in the Brillouin zone (shown in Fig. 2b). We find that the conduction bands are mostly composed of the hybridized Bi-6$p$ and In-5$s$ orbitals, whereas the valence bands are dominated by the O-2$p$ orbitals with a small admixture of the Bi-6$s$ states. It is seen from Fig. 4a that BiInO$_3$ is an indirect band-gap semiconductor with the valence band maximum (VBM) located at the T point and the conduction band minimum (CBM) located along the Γ-X symmetry line. The calculated band gap is about 2.6 eV.

Including SOC (Fig. 4b) reduces the band gap to about 2.3 eV and strongly affects the electronic structure of conduction bands of BiInO$_3$. Comparing the band structures calculated with SOC (Fig. 4b) and without SOC (Fig. 4a), a sizable band spin splitting produced by the SOC is seen at some high symmetry **k** points and along certain **k** paths. At the X point, which is located in the proximity of the CBM, the two lower energy states are doublets resulting from the SOC splitting. The splitting is large, i.e. Δ ≈ 0.26 eV. As expected, the bands along the X-S line are double degenerate protected by the Θ symmetry. When moving from the X to Γ point the doublets are split into singlets with a nearly linear dispersion (see inset of Fig. 4b).

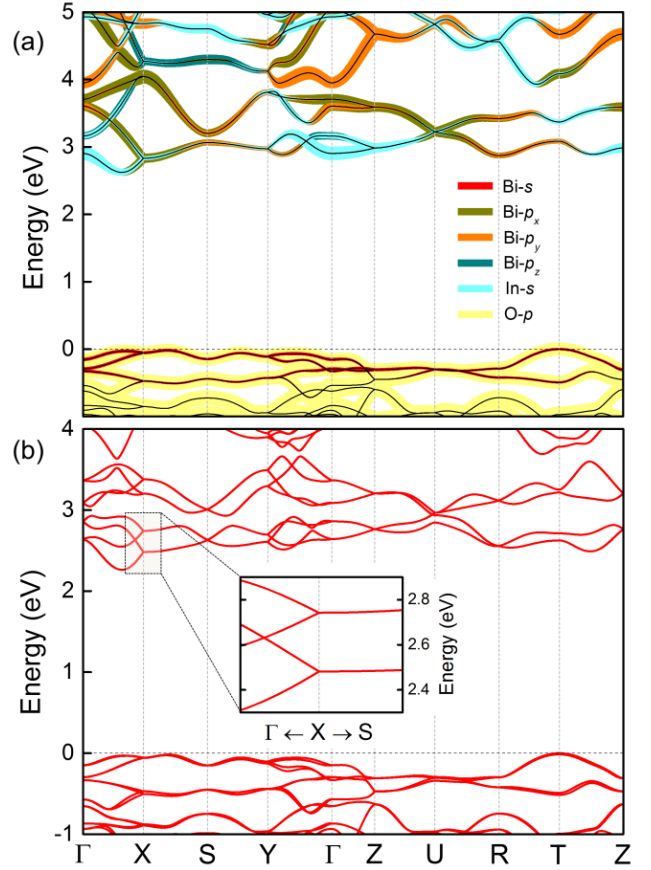

**Fig. 4 Band structure of bulk BiInO$_3$. (a, b)** Band structure along the high symmetry lines in the Brillouin zone without SOC (a) and with SOC (b). Orbital-contributions in panel (a) are shown by color lines with thickness proportional to the orbital weight. Inset in panel (b) shows the band structure zoomed in around the X point.

Lifting the degeneracy along the Γ-X line, where $\mathbf{k} = (k_x, 0, 0)$, can be understood from the little group of wave vector **k**, which has symmetry generators $\bar{M}_y$ and $\tilde{\Theta} \equiv T\bar{M}_x$. Along this line $\tilde{\Theta}^2 = T^2 \bar{M}_x^2 = e^{-ik_y - ik_z} = 1$ and thus the Bloch states $\psi_\mathbf{k}$ and $\tilde{\Theta}\psi_\mathbf{k}$ are not degenerate. In addition, each state $\psi_\mathbf{k}$ can be labeled using the eigenvalues of $\bar{M}_y$. Since $\bar{M}_y^2 = -e^{-ik_x}$, we obtain $\bar{M}_y |\psi_\mathbf{k}^\pm\rangle = \pm i e^{-i\frac{k_x}{2}} |\psi_\mathbf{k}^\pm\rangle$. Therefore, there are four non-degenerate Bloch states, $\psi_{\mathbf{k}}^{\pm}$ and $\tilde{\Theta}\psi_{\mathbf{k}}^{\pm}$, evolving from the X point when moving along the X-Γ line (inset of Fig. 4b). Interestingly, crossing the $\psi_{\mathbf{k}}^{+}$ and $\psi_{\mathbf{k}}^{-}$ bands is enforced and protected by symmetry, resulting in a hourglass-shaped band dispersion[24,25] (see Supplementary Note 1).

The SOC splitting at the Y point is smaller Δ ≈ 0.09 eV. The bands along the Y-S line are double degenerate protected by the $\tilde{\Theta}$ symmetry, but split when moving from the Y to Γ point. This behavior can be understood using the considerations similar to those we used to explain band degeneracies and splittings around the X point.

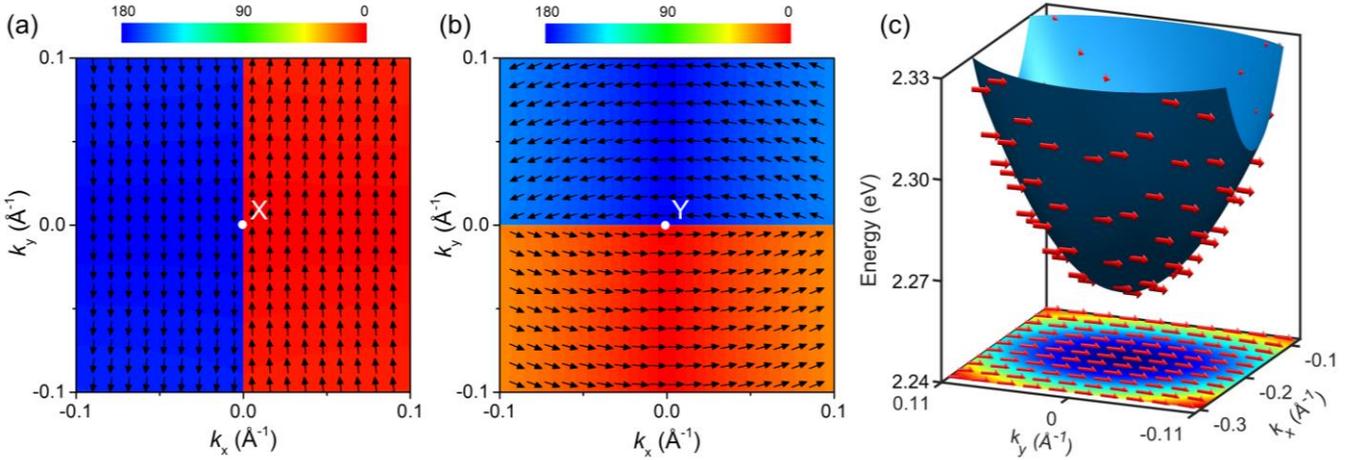

**Fig. 5 Spin texture of BiInO$_3$. (a, b)** Spin configurations around the high-symmetry **k** points: X point (a) and Y point (b). The spin textures are plotted in the $k_z = 0$ plane for the lowest energy conduction bands. The wave vector **k** is referenced to the X point (a) and Y point (b) where it is assumed to be zero. The color map reflects the polar angle (in degrees) with respect to the *y* axis (a) and *x* axis (b). **(c)** 3D diagram and 2D projection of band structure and spin texture around the CBM. The arrows indicate the spin direction. The color map shows the energy profile. The wave vector is referenced to the X point where is it assumed to be zero.

Next we explore the spin texture around the X and Y points. According to Table 1, the space group No. 33 for BiInO$_3$ supports the PST with the uniform spin orientation along the *y* (*x*) axis around the X (Y) point. This is exactly what we find from our DFT results. Fig. 5a shows the calculated spin texture around the X point for the conduction band, which has the lowest energy. We see a unidirectional spin configuration for positive and negative values of $k_x$ (referred to the origin being at the X point), which is consistent with the effective SOC field $\mathbf{\Omega}_{\text{PST}} = \lambda(0, k_x, 0)$ and the PST in Fig. 1. As expected, the spin orientation changes abruptly at $k_x = 0$, where the $s_y$ component of the spin is reversed. We note that there is another band (with the opposite $\bar{M}_y$ eigenvalue if moving along the X-Γ line) which has higher energy (except the X-S line where it has the same energy) for which the spin has opposite orientation.

It is remarkable that the PST covers a substantial part of the Brillouin zone. The range of $k_x$ and $k_y$ values in Fig. 5a spans 0.2 Å$^{-1}$ around the X point. For comparison, the *x* component of the reciprocal wave vector is $\pi/a = 0.528$ Å$^{-1}$ (distance from the to X to Γ point). In fact, the nearly uniform spin structure persists even at larger distances and covers the CBM which is located at about 0.19 Å$^{-1}$ from the X point along the X-Γ line. Fig. 5c shows the band dispersion and the spin structure around the CBM. It is evident that the spin maintains nearly unidirectional texture along the *y* direction, which is reminiscent to that at the X point. Our calculations predict that in the range of $k_x$ and $k_y$ values spanning 0.2 Å$^{-1}$ around the CBM (as in Fig. 5c), the largest deviation of the spin orientation from the *y* axis is only 9.6°. This is due the CBM-forming band being well separated from the other two bands derived from the higher energy doublet (inset of Fig. 4b), so that the mixing between the doublets is minor. We note that the PST around the CBM is fully reversed when the wave vector **k** is changed to –**k**, due to time reversal symmetry.

The spin structure around the Y point (Fig. 5b) shows the similar trend, now with the spin being textured along the *x* direction. The effective SOC field $\mathbf{\Omega}_{\text{PST}} = \lambda(k_y, 0, 0)$ in this case leads to reversal of the $s_x$ component when crossing the $k_y = 0$ line. There is a visible deviation from the unidirectional spin orientation when moving far away from the Y point. This stems from the reduced SOC splitting at the Y point (Δ ≈ 0.09 eV) as compared to that the X point (Δ ≈ 0.26 eV).

**A k·p model.** The spin textures around the high symmetry points can be further understood in terms of an effective **k·p** Hamiltonian, which we deduce from symmetry considerations. Here, we focus on the X point. In order to describe the four dispersing bands around the X point, additional sublattice degrees of freedom need to be included in the consideration. These are conventionally described by a set of Pauli matrices $\tau_j$ ($j = x, y, z$) operating in the sublattice space. The Hamiltonian around the X point is constructed by taking into account all symmetry operations at the X point, at which the symmetry generators are $\bar{M}_x$ and $\bar{M}_y$, and the time-reversal symmetry, which operator is $T = i\sigma_y K$, where $K$ is complex conjugation. We find that $\bar{M}_x$ and $\bar{M}_y$ can be represented as $\bar{M}_x = i\tau_z\sigma_x$ and $\bar{M}_y = \tau_y\sigma_y$ (see Supplementary Note 2). Collecting all the terms up to linear order in **k**, which are invariant under these symmetry transformations, we obtain the $\mathbf{k}\cdot\mathbf{p}$ Hamiltonian:

$$H = \delta\tau_y\sigma_y + \alpha k_x\tau_0\sigma_y + \beta k_y\tau_0\sigma_x + \gamma_1 k_x\tau_y\sigma_0 + \gamma_2 k_x\tau_x\sigma_x + \gamma_3 k_x\tau_z\sigma_z + \gamma_4 k_y\tau_x\sigma_y. \quad (5)$$

Here for simplicity we limit our consideration by the $(k_x, k_y)$ plane; $\delta$, $\alpha$, $\beta$, $\gamma_m$ ($m = 1$–4) are independent parameters, $\sigma_0$ and $\tau_0$ are the 2×2 identity matrices, and direct products $\tau_i \otimes \sigma_j$ ($i, j = 0, x, y, z$) are implicitly assumed.

When $k = 0$ (i.e. at the X point), the $\delta\tau_y\sigma_y$ term in the Hamiltonian splits the state into two doublets distinguished by the eigenvalues of $\bar{M}_y = \pm 1$ and separated by $\Delta = 2\delta$. When $k$ is not too large, the other terms in the Eq. (5) can be treated as perturbation. In the first order, the perturbation does not mix the doublets and the effective Hamiltonian within each of the doublets (labelled by indices $\pm$) can be written as

$$H^\pm(k_x) = \pm\delta + \alpha^\pm k_x\sigma_y, \quad (6)$$

where $\alpha^\pm = \sqrt{(\alpha \pm \gamma_1)^2 + (\gamma_2 \mp \gamma_3)^2}$ (see Supplementary Note 2). The corresponding eigenvalues are

$$\begin{aligned} E^+(k_x) &= \delta \pm \alpha^+ k_x \\ E^-(k_x) &= -\delta \pm \alpha^- k_x, \end{aligned} \quad (7)$$

i.e. each doublet is split into two singlet states exhibiting linear dispersion away from the X point. This is consistent with our DFT results (inset in Fig. 4b). Fitting the DFT energy bands yields the following parameters: $\delta = -0.13$ eV, $\alpha^+ = 1.91$ eV Å, $\alpha^- = 1.51$ eV Å. Other parameters in the Hamiltonian of Eq. (5), can be found from the expectation values of the $y$-component of the spin, $s_y = \pm\frac{1}{2}(\alpha + \gamma_1)/\alpha^+$ and $s_y = \pm\frac{1}{2}(\alpha - \gamma_1)/\alpha^-$, for the doublet (+) and doublet (−), respectively. Using the DFT results for $s_y$, we obtain $\alpha = -0.18$ eV Å, $\gamma_1 = -1.42$ eV Å, $\gamma_2 = 0.95$ eV Å, and $\gamma_3 = 0.09$ eV Å.

The effective Hamiltonian of Eq. (6) imposes the effective SOC field pointing along the $y$ direction, i.e. $\mathbf{\Omega}_{PST} = \lambda(0, k_x, 0)$, where $\lambda = \alpha^\pm$, which produces the PST (Fig. 1c). Importantly, this form of the PST Hamiltonian appears as the result of the crystal symmetry rather than matching the Rashba and Dresselhaus constants. For the lowest energy band, $\lambda = \alpha^+$ and the spin is parallel (antiparallel) to the $y$-direction for positive (negative) $k_x$. This is in agreement with the spin structure in Fig. 5a obtained from our DFT calculations.

Including first-order perturbation corrections to the wave function mixes states between the doublets, resulting in non-vanishing components of $s_x$ and $s_z$ and thus deviation from the PST. Our detailed analysis (see Supplementary Note 3) shows that within this approximation the $s_y$ component of the spin remains constant (Supplementary Eq. 21), whereas the $s_x$ component varies as $s_x = qk_y/\Delta$ (for the lowest conduction band), where $q$ is a SOC constant (Supplementary Eq. 22). It is evident from this result that, first, $s_x = 0$ at $k_y = 0$ and hence the spin orientation remains collinear to the $y$ axis at the CBM ($k_x = 0.19$ Å$^{-1}$, $k_y = 0$) as at the X point, and, second, when going away from the CBM along $k_y$ the $s_x$ value changes linear with $k_y$. Non-zero $s_x$ produces deviation from PST, but this deviation remains small over a broad area around the CBM due to the large splitting $\Delta$. This is evident from Supplementary Fig. 3 which also reveals excellent agreement between the perturbation theory and explicit DFT calculation. This approach also allows us to obtain the remaining SOC constants in Hamiltonian of Eq. (5), $\beta = -0.139$ eV Å and $\gamma_4 = 0.191$ eV Å, as detailed in Supplementary Note 3.

We would like to note that for the compounds considered in our work, any order terms in $k$ in the Hamiltonian preserve PST in zero-order perturbation theory. Only in the first order of perturbation theory for the wave function a deviation from PST occurs with the dominant contribution resulting from linear in $k$ terms. However, due to this contribution occurring as a perturbation, deviation from the PST remains small over a large area of the Brillouin zone.

## Discussion

The obtained value of the SOC parameter $\lambda = 1.91$ eV Å is three orders of magnitude larger than the values known for the semiconductor quantum-well structures (1–5 meV Å).[15-18] It is also larger than the values predicted for other ferroelectric oxides, e.g., $\lambda_R = 0.74$ eV Å for BiAlO$_3$ (*P4mm* space group)[26] and $\lambda_D = 0.58$ eV Å for HfO$_2$ (*Pca*2$_1$ space group),[27] and comparable to the value of $\lambda_R = 3.85$ eV Å observed in BiTeI.[28] The associated band splittings are sufficient to support room temperature functionalities. For example, the lowest excited state at $\mathbf{k} \approx (0.64\pi, 0, 0)$ corresponding to the CBM lies about 0.29 eV above the CBM.

Electron motion in the PST state forms persistent spin helix (PSH) – the spatially periodic mode of spin polarization with the wave length of $l_{PSH} = \frac{\pi \hbar^2}{m\lambda}$.[13] We estimate the effective mass $m$ in BiInO$_3$ by fitting the band dispersion around the CBM, which leads to $m = 0.61 m_0$, where $m_0$ is the free electron mass. The resulting wave length is about 2 nm. This value is three orders of magnitude smaller than $l_{PSH} \sim 5-10\,\mu$m observed in semiconductor heterostructures.[16]

It is conceivable (though challenging) to form and map a PSH state in BiInO$_3$ in spirit of experiments by Walser et al.[16] BiInO$_3$ is a wide band gap semiconductor, and in order to observe this property an electron doping is required. Since Bi is isovalent to In, In$_2$O$_3$ may be considered as a comparative compound. It is known that oxygen vacancies naturally form in In$_2$O$_3$ producing $n$-type conductivity which can be varied over a broad range of magnitudes by changing growth conditions (mainly oxygen pressure).[29] We expect, therefore, that a similar approach could be employed to produce electron doping in BiInO$_3$. Due to CBM in BiInO$_3$ maintaining PST, a PSH state will be formed if electrons are optically injected into the conduction band of BiInO$_3$. Mapping the formation and evolution of PSH in BiInO$_3$ could possibly be performed using near-field scanning Kerr microscopy, which showed a possibility to resolve features down to tens-nm scale with sub-ns time resolution.[30] In addition, the electron-doped BiInO$_3$ can be used to explore the current induced spin polarization (known as the Edelstein effect[7]) and associated spin-orbit torques,[31] which are expected to be large due to the large SOC.

We also envision a possibility to observe a Hall effect qualitatively similar to the valley Hall effect recently discovered in transition metal dichalcogenides (TMD).[32] In BiInO$_3$ the two states with $\mathbf{k}$ and -$\mathbf{k}$ at the CBM with opposite spin orientation are related by time reversal symmetry transformation and thus have opposite sign of the Berry curvature. If an imbalance in electron population between these two states is created by polarized optical excitation (similar to that done in TMDs), a charge Hall current can be measured that reverses sign with polarization of the exciting light.

Another implication is a possibility to use a PST material as a barrier in tunnel junctions. It has been predicted that the Rashba and Dresselhaus SOC in a tunnel barrier can produce tunneling anomalous and spin Hall effects.[33,34] Using a PST material as a tunnel barrier allows producing a perfect anisotropy in the Hall response. For example, if the current flows in the $z$ direction across a PST barrier with the SOC given by Eq. (6), the tunneling Hall response will be zero in the $y$ direction and non-zero in the $x$ direction. Moreover, the anomalous Hall conductivity is expected to strongly depend on the magnetization orientation in the $x$-$y$ plane and vanish for magnetization pointing along the $x$ direction. The large value of $\lambda = 1.91$ eV Å is expected to produce sizable effects, which can be detected experimentally. In addition, the reversible spin texture of ferroelectric SOC oxide materials[35,36] will support the tunneling Hall effects to be reversible by an applied electric field through switching of ferroelectric polarization.[27]

Apart from BiInO$_3$, there are a number of other potential candidates which are expected to maintain a PST. Among them are BiInS$_3$ (*Pna*2$_1$ structure, space group No. 33) and LiTeO$_3$ (*Pnn*2 structure, space group No. 34). Both have a PST around the high symmetry X and Y points (see Supplementary Note 4). BiInS$_3$ has a lower calculated band gap (about 1.15 eV), but a CBM is located at the Γ point which does not support the PST. On the other hand, LiTeO$_3$ (calculated band gap is about 2 eV) has a CBM close to the X point similar to BiInO$_3$.

Overall, we have demonstrated that the PST is imposed by symmetry in a class of orthorhombic non-symmorphic bulk materials, such as BiInO$_3$. The PST is a robust intrinsic property of these materials, which eliminates the stringent condition of equal Rashba and Dresselhaus SOC for realizing the persistent spin helix. The electronic and spin properties of the PST materials are derived from the non-trivial interplay between spin-orbit coupling and glide reflection symmetries, and in this regard place them among interesting quantum materials which have recently received a lot of attention. We hope therefore that our theoretical predictions will stimulate experimental efforts in the exploration of these materials, which functional properties may be useful for device applications.

## Methods

**DFT calculations**. DFT calculations are performed using a plane-wave pseudopotential method implemented in Quantum-ESPRESSO.[37] In the calculations, we use the lattice constants and atomic positions of bulk materials, which are given in Supplementary Note 4. The exchange-correlation functional is treated within the generalized gradient approximation (GGA).[38] We use energy cutoff of 544 eV for the plane wave expansion and 10×10×8 $k$-point grid for Brillouin zone integrations. The electric polarization is computed using the Berry phase method.[39] SOC is included in the calculations using the fully-relativistic ultrasoft pseudopotentials.[40] The expectation values of the spin operators $s_i = \frac{1}{2}\langle \psi_{\mathbf{k}} | \sigma_i | \psi_{\mathbf{k}} \rangle$ ($i = x, y, z$) are obtained directly from the non-collinear spin DFT calculations. The atomic structures are produced using VESTA software.[41]

**Data availability.** The data that support the findings of this study are available from the authors upon request.

**Acknowledgments**

This work was supported by the National Science Foundation (NSF) through Nebraska Materials Research Science and Engineering Center (MRSEC) (Grant No. DMR-1420645). Computations were performed at the University of Nebraska Holland Computing Center. The authors thank Jaroslav Fabian, Jairo Sinova, Silvia Picozzi, and Alexei Kovalev for helpful discussions.

**Author contributions**

L.L.T. and E.Y.T conceived the project. L.L.T. carried out DFT calculations. L.L.T. and E.Y.T. performed the symmetry analysis and theoretical modeling. Both authors discussed the results and wrote the manuscript.

**Competing interests**: The authors declare no competing interests.

# Supplementary Information

## Persistent spin texture enforced by symmetry

Tao *et al.*

## Supplementary Note 1.

**Projected density of states, hourglass band dispersion, and a nodal line in BiInO$_3$**

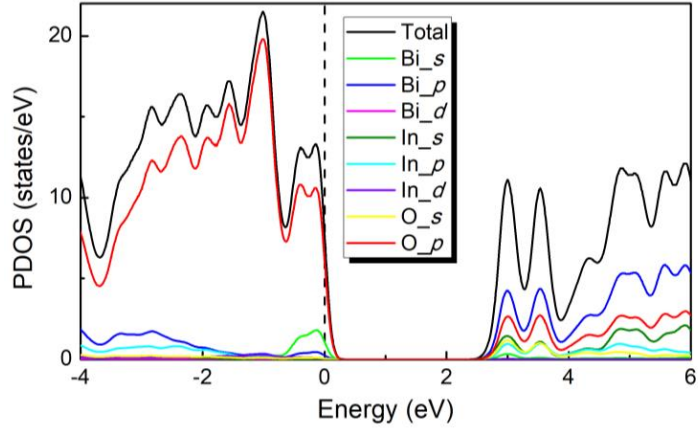

**Supplementary Figure 1. Projected density of states**. Projected density of states (PDOS) for bulk BiInO$_3$ without SOC.

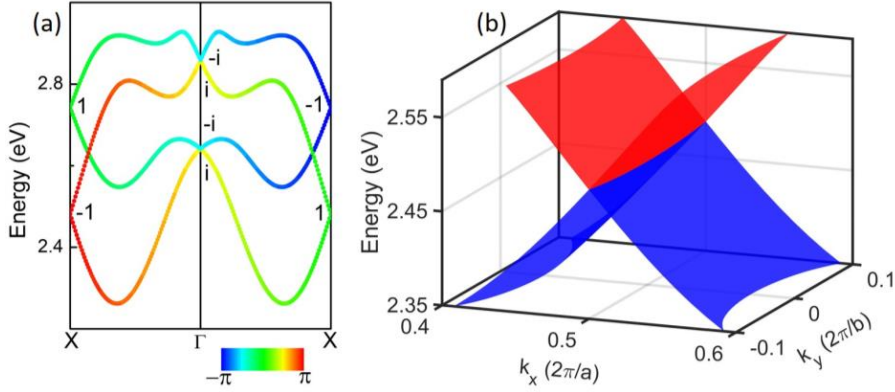

**Supplementary Figure 2. Hourglass-shaped band dispersion and node line.** (a) Lowest energy conduction bands along the Γ-X symmetry line. The color map quantifies the phase of the eigenvalues of $\bar{M}_y$. The numbers represent the eigenvalues at the time-reversal-invariant momenta, i.e. Γ and X. (b) Three-dimensional band dispersion for the lowest two conduction bands around the X point.

As discussed in the main text, along the Γ-X line, where $\mathbf{k} = (k_x, 0, 0)$, $\mathbf{k}$ is invariant under $\bar{M}_y$ operation. Note that transformation rule for wave vector $\mathbf{k}$ ($k_x$, $k_y$, $k_z$) under the glide reflection is the same as that under the mirror reflection. Each Bloch state $\psi_\mathbf{k}$ can be labeled using the eigenvalues of $\bar{M}_y$: $\bar{M}_y |\psi_\mathbf{k}^\pm\rangle = \pm i e^{-i\frac{k_x}{2}} |\psi_\mathbf{k}^\pm\rangle$. In addition, along this line, $\tilde{\Theta}^2 = T^2 \bar{M}_x^2 = e^{-ik_y - ik_z} = 1$, and thus $\tilde{\Theta}$ cannot result in the Kramers-like double degeneracy. $T$ enforces the band degeneracy at the time-reversal-invariant momenta (TRIMs) $\mathbf{k}$, which satisfy $\mathbf{k} = -\mathbf{k} + \mathbf{G}$ with $\mathbf{G}$ being the reciprocal lattice vector. At the TRIMs, the Kramers' pair has the complex-conjugate eigenvalues of $\bar{M}_y$, namely (i, – i) and (–i, i) at the Γ point. In contrast, the eigenvalues for Kramers' pair are (1, 1) and (–1, –1) at the X point. The eigenvalue of $\bar{M}_y$ must evolve continuously from i to 1 or from -i to –1, resulting in an hourglass-shaped band dispersion (Supplementary Fig. 2a).[1,2] The band crossing occurs between the Γ and X point, which is protected by the $\bar{M}_y$ symmetry and robust against the SOC. This is seen from the following consideration. Assume that there exists perturbation $H'$ in the system. Since $\bar{M}_y |\psi_{\mathbf{k}}^{\pm}\rangle = \pm i e^{-i\frac{k_x}{2}} |\psi_{\mathbf{k}}^{\pm}\rangle$, the matrix element of $H'$ is given by $\langle\psi_{\mathbf{k}}^{+}|H'|\psi_{\mathbf{k}}^{-}\rangle = \langle\psi_{\mathbf{k}}^{+}|\bar{M}_y^{-1}H'\bar{M}_y|\psi_{\mathbf{k}}^{-}\rangle = -\langle\psi_{\mathbf{k}}^{+}|H'|\psi_{\mathbf{k}}^{-}\rangle$, which implies $\langle\psi_{\mathbf{k}}^{+}|H'|\psi_{\mathbf{k}}^{-}\rangle = 0$. Thus, any perturbation that does not break the $\bar{M}_y$ symmetry cannot induce hybridization between the states with different $\bar{M}_y$ eigenvalues. The crossing point is therefore protected by the $\bar{M}_y$ symmetry. On the other hand, while the bands are split along the Γ-X line, they are double degenerate along the X-S line. Therefore, a node line along $\mathbf{k} = (\pm\pi, k_y, 0)$ is naturally formed (Supplementary Fig. 2b).

## Supplementary Note 2.
### A model Hamiltonian

Here, we derive the $\mathbf{k} \cdot \mathbf{p}$ effective Hamiltonian for BiInO$_3$ around the X point and determine the eigenvalues and eigenstates using the perturbation theory. The $\mathbf{k} \cdot \mathbf{p}$ Hamiltonian around the X point can be constructed by considering all the symmetry operations at the X point[3], at which the symmetry generators are $\bar{M}_x$ and $\bar{M}_y$. As shown in the main text, there are two conjugated doublets at the X point, $(\psi_X^+, \Theta\psi_X^+)$ or $(\psi_X^-, \Theta\psi_X^-)$, which are distinguished by the $\bar{M}_y$ eigenvalues and $\Theta \equiv T\bar{M}_y$. To describe these four states, in addition to spin, sublattice degrees of freedom need to be included in the consideration, which are conventionally described by a set of Pauli matrices $\tau_j$ ($j = x, y, z$). The time-reversal symmetry $T$ is represented as $T = i\sigma_y K$, where $K$ is complex conjugation. The $\bar{M}_y$ operator in the spin space is described by $i\sigma_y$. In order to determine its effect on the pseudospin $\mathbf{\tau}$, we take into account the fact that $\bar{M}_y^2 = 1$ at the X point and that $[T, \bar{M}_y] = 0$. This leads to $\bar{M}_y = \tau_y \sigma_y$. In order to determine $\bar{M}_x$, we calculate the eigenvalues of $\bar{M}_x$ for the doublets $(\psi_X^+, \Theta\psi_X^+)$ or $(\psi_X^-, \Theta\psi_X^-)$. Since $\bar{M}_x^2 = -e^{-ik_y} = -1$ at the X point, we have $\bar{M}_x |\psi_X^{\pm}\rangle = \pm i |\psi_X^{\pm}\rangle$. Now we calculate the eigenvalue of $\bar{M}_x$ for the conjugated state $\Theta|\psi_X^{\pm}\rangle$. The commutation relation between $\bar{M}_x$ and $\bar{M}_y$ can be derived from the following successive symmetry operations. In real space, we obtain

$$\left.\begin{array}{l}(x,y,z) \xrightarrow{\bar{M}_x} (-x+\tfrac{1}{2}, y+\tfrac{1}{2}, z+\tfrac{1}{2}) \xrightarrow{\bar{M}_y} (-x+1, -y, z+\tfrac{1}{2}) \\ (x,y,z) \xrightarrow{\bar{M}_y} (x+\tfrac{1}{2}, -y+\tfrac{1}{2}, z) \xrightarrow{\bar{M}_x} (-x, -y+1, z+\tfrac{1}{2})\end{array}\right\} \Rightarrow \bar{M}_x \bar{M}_y = e^{-ik_x + ik_y} \bar{M}_y \bar{M}_x. \quad (1)$$

In the spin space, $\bar{M}_x \bar{M}_y = -\bar{M}_y \bar{M}_x$ due to $\bar{M}_x = i\sigma_x$ and $\bar{M}_y = i\sigma_y$. Combining the real space and spin space, we obtain $\bar{M}_x \bar{M}_y = -e^{-ik_x + ik_y} \bar{M}_y \bar{M}_x$. At the X point ($k_x = \pi$, $k_y = 0$), we then have $[\bar{M}_x, \bar{M}_y] = 0$. Thus, $\bar{M}_x \Theta|\psi_X^{\pm}\rangle = \Theta\bar{M}_x|\psi_X^{\pm}\rangle = \mp i\Theta|\psi_X^{\pm}\rangle$, implying that the eigenvalues for the doublets $(\psi_X^+, \Theta\psi_X^+)$ or $(\psi_X^-, \Theta\psi_X^-)$ have opposite sign. To satisfy all these conditions $\bar{M}_x$ can be chosen as $\bar{M}_x = i\tau_z \sigma_x$. From the expressions for $T$, $\bar{M}_x$, and $\bar{M}_y$ given above, it is easy to see that $[T, \bar{M}_x] = 0$ and $[T, \bar{M}_y] = 0$ in the spin space. Since $T$ does not change the real-space coordinates, the commutation relations $[T, \bar{M}_x] = 0$ and $[T, \bar{M}_y] = 0$ also hold in the real space. The corresponding transformations for $\mathbf{k}$, $\mathbf{\sigma}$ and $\mathbf{\tau}$ are given in Supplementary Table 1.

**Supplementary Table 1.** Transformation rules for wave vector **k**, and spin ($\sigma$) and sublattice ($\tau$) Pauli matrices under the $C_{2v}$ point-group symmetry operations at the X ($\pi$, 0, 0) point in the Brillouin zone of BiInO$_3$. The wave vector **k** is referenced with respect to the high symmetry point where it is assumed to be zero. K denotes complex conjugation.

| Symmetry | $(k_x, k_y, k_z)$ | $(\sigma_x, \sigma_y, \sigma_z)$ | $(\tau_x, \tau_y, \tau_z)$ |
|---|---|---|---|
| $T = i\sigma_y K$ | $(-k_x, -k_y, -k_z)$ | $(-\sigma_x, -\sigma_y, -\sigma_z)$ | $(\tau_x, -\tau_y, \tau_z)$ |
| $\bar{M}_x = i\tau_z \sigma_x$ | $(-k_x, k_y, k_z)$ | $(\sigma_x, -\sigma_y, -\sigma_z)$ | $(-\tau_x, -\tau_y, \tau_z)$ |
| $\bar{M}_y = \tau_y \sigma_y$ | $(k_x, -k_y, k_z)$ | $(-\sigma_x, \sigma_y, -\sigma_z)$ | $(-\tau_x, \tau_y, -\tau_z)$ |
| Commutation | $[T, \bar{M}_x] = 0$, $[T, \bar{M}_y] = 0$, $[\bar{M}_x, \bar{M}_y] = 0$, $\bar{M}_x^2 = -1$, $\bar{M}_y^2 = 1$ | | |

We limit our consideration by dispersion in the $(k_x, k_y)$ plane. Collecting all the terms up to linear order in **k** (the quadric and cubic in $k$ terms are listed in Supplementary Table 3 of Supplementary Note 6), which are invariant under these symmetry transformations, we obtain the $\mathbf{k} \cdot \mathbf{p}$ Hamiltonian as follows:

$$H = \delta \tau_y \sigma_y + \alpha k_x \tau_0 \sigma_y + \beta k_y \tau_0 \sigma_x + \gamma_1 k_x \tau_y \sigma_0 + \gamma_2 k_x \tau_x \sigma_x + \gamma_3 k_x \tau_z \sigma_z + \gamma_4 k_y \tau_x \sigma_y , \tag{2}$$

where **k** is referred to the X point. We split the Hamiltonian of Eq. (2) into $H = H_0 + H'$, where

$$H_0 = \delta \tau_y \sigma_y \tag{3}$$

is $k$ independent term and

$$H' = \alpha k_x \tau_0 \sigma_y + \beta k_y \tau_0 \sigma_x + \gamma_1 k_x \tau_y \sigma_0 + \gamma_2 k_x \tau_x \sigma_x + \gamma_3 k_x \tau_z \sigma_z + \gamma_4 k_y \tau_x \sigma_y \tag{4}$$

is $k$ dependent term. $H'$ can be treated as perturbation when $k$ is small (measured from the X point). The basis set can be constructed as the direct product of the eigenstates for pseudospin $\tau$ and spin $\sigma$. It is convenient to choose basis functions to be the eigenstates of $\tau_y$ and $\sigma_y$. We use $\chi$ and $\xi$ to denote the spin eigenstates for $\tau_y$ and $\sigma_y$, respectively, so that $\chi_+ = \frac{1}{\sqrt{2}}\begin{bmatrix}1\\i\end{bmatrix}$, $\chi_- = \frac{1}{\sqrt{2}}\begin{bmatrix}i\\1\end{bmatrix}$ and $\xi_+ = \frac{1}{\sqrt{2}}\begin{bmatrix}1\\i\end{bmatrix}$, $\xi_- = \frac{1}{\sqrt{2}}\begin{bmatrix}i\\1\end{bmatrix}$. Then the basis set is as follows

$$\begin{cases} \phi_1 = \chi_+ \xi_+ \\ \phi_2 = \chi_- \xi_- \\ \phi_3 = \chi_+ \xi_- \\ \phi_4 = \chi_- \xi_+ \end{cases} . \tag{5}$$

Hamiltonian $H_0$ in the basis set is diagonal, so that $H_0 = \text{diag}[\delta, \delta, -\delta, -\delta]$, and its eigenvalues are

$$\begin{cases} E_{1,2}^{(0)} = \delta \\ E_{3,4}^{(0)} = -\delta \end{cases} . \tag{6}$$

In the basis of $\phi_j$ ($j$ = 1–4), the matrix elements of the spin operators $\sigma_j$ ($j = x, y, z$) can be expressed as

$$\sigma_x = \begin{bmatrix} 0 & 0 & 1 & 0 \\ 0 & 0 & 0 & 1 \\ 1 & 0 & 0 & 0 \\ 0 & 1 & 0 & 0 \end{bmatrix}, \quad \sigma_y = \begin{bmatrix} 1 & 0 & 0 & 0 \\ 0 & -1 & 0 & 0 \\ 0 & 0 & -1 & 0 \\ 0 & 0 & 0 & 1 \end{bmatrix}, \quad \sigma_z = \begin{bmatrix} 0 & 0 & i & 0 \\ 0 & 0 & 0 & -i \\ -i & 0 & 0 & 0 \\ 0 & i & 0 & 0 \end{bmatrix}. \tag{7}$$

We see that the eigenstates of the Hamiltonian $H_0$ diagonalize the $\sigma_y$ matrix (i.e. are eigenstates of $\sigma_y$), which is consistent with the symmetry arguments of the main text.

Next, we follow the standard degenerate perturbation theory to determine the eigenvalues and eigenstates. In the basis of $\phi_j$ ($j = 1$–4), $H'$ can be expressed as

$$H' = \begin{bmatrix} \alpha k_x + \gamma_1 k_x & \gamma_2 k_x - \gamma_3 k_x & \beta k_y & \gamma_4 k_y \\ \gamma_2 k_x - \gamma_3 k_x & -\alpha k_x - \gamma_1 k_x & -\gamma_4 k_y & \beta k_y \\ \beta k_y & -\gamma_4 k_y & -\alpha k_x + \gamma_1 k_x & \gamma_2 k_x + \gamma_3 k_x \\ \gamma_4 k_y & \beta k_y & \gamma_2 k_x + \gamma_3 k_x & \alpha k_x - \gamma_1 k_x \end{bmatrix}. \tag{8}$$

The first order correction to the energy $E^{(1)}$ is

$$\begin{cases} E_1^{(1)} = -\alpha^+ k_x, & E_2^{(1)} = \alpha^+ k_x \\ E_3^{(1)} = -\alpha^- k_x, & E_4^{(1)} = \alpha^- k_x \end{cases}, \tag{9}$$

where

$$\alpha^\pm = \sqrt{(\alpha \pm \gamma_1)^2 + (\gamma_2 \mp \gamma_3)^2}, \tag{10}$$

and the eigenstates $\psi^{(0)}$ are

$$\begin{cases} \psi_1^{(0)} = \dfrac{1}{\sqrt{c_1^2+1}}(\phi_1 + c_1\phi_2), & \psi_2^{(0)} = \dfrac{1}{\sqrt{c_1^2+1}}(-c_1\phi_1 + \phi_2) \\ \psi_3^{(0)} = \dfrac{1}{\sqrt{c_2^2+1}}(\phi_3 - c_2\phi_4), & \psi_4^{(0)} = \dfrac{1}{\sqrt{c_2^2+1}}(c_2\phi_3 + \phi_4) \end{cases}, \tag{11}$$

where the parameters $c_j$ ($j = 1$–2) are defined as

$$\begin{cases} c_1 = \dfrac{\gamma_2 - \gamma_3}{\alpha + \gamma_1 - \alpha^+} \\ c_2 = \dfrac{\gamma_2 + \gamma_3}{\alpha - \gamma_1 + \alpha^-} \end{cases}. \tag{12}$$

Within the basis set of functions (Supplementary Eq. (11)) the matrix elements of $\sigma_j$ ($j = x, y, z$) are given by

$$\sigma_x = \frac{1}{\sqrt{(c_1^2+1)(c_2^2+1)}} \begin{bmatrix} 0 & 0 & 1-c_1c_2 & c_1+c_2 \\ 0 & 0 & -c_1-c_2 & 1-c_1c_2 \\ 1-c_1c_2 & -c_1-c_2 & 0 & 0 \\ c_1+c_2 & 1-c_1c_2 & 0 & 0 \end{bmatrix}$$

$$\sigma_y = \begin{bmatrix} \frac{1-c_1^2}{c_1^2+1} & 0 & 0 & 0 \\ 0 & \frac{c_1^2-1}{c_1^2+1} & 0 & 0 \\ 0 & 0 & \frac{c_2^2-1}{c_2^2+1} & 0 \\ 0 & 0 & 0 & \frac{1-c_2^2}{c_2^2+1} \end{bmatrix}$$

$$\sigma_z = \frac{i}{\sqrt{(c_1^2+1)(c_2^2+1)}} \begin{bmatrix} 0 & 0 & 1+c_1c_2 & c_2-c_1 \\ 0 & 0 & c_2-c_1 & -1-c_1c_2 \\ -1-c_1c_2 & c_1-c_2 & 0 & 0 \\ c_1-c_2 & 1+c_1c_2 & 0 & 0 \end{bmatrix}. \tag{13}$$

We see that in agreement with our general consideration given in the main text, the matrix elements of the spin operators $\sigma_x$ and $\sigma_z$ within the two doublets are equal to zero (two 2×2 block diagonal matrices). On the other hand, we also see that $\psi_j^{(0)}$ ($j=1$–$4$) are the eigenstates of $\sigma_y$ and have eigenvalues of opposite sign for the two states within either doublet. This allows us writing an effective Hamiltonian up to the first order in perturbation theory within each of the two doublets (labelled by indices $\pm$) in the form of in Eq. (6) of the main text, i.e.

$$H^\pm(k_x) = \pm\delta + \alpha^\pm k_x \sigma_y. \tag{14}$$

We note that, the expectation values of $s_y = \frac{1}{2}\langle\sigma_y\rangle$ are not equal to $\pm\frac{1}{2}$ but in general deviate from these values.[4] Using Supplementary Eqs. (12) and (13), we find

$$s_y = \pm\frac{1}{2}\frac{\alpha+\gamma_1}{\alpha^+} \tag{15}$$

and

$$s_y = \pm\frac{1}{2}\frac{\alpha-\gamma_1}{\alpha^-} \tag{16}$$

for doublet (+) and doublet (−), respectively.

These relationships can be used to find all the parameters in the model Hamiltonian Supplementary Eq. (2). The values of $\delta$, $\alpha^+$, and $\alpha^-$ are obtained by fitting the DFT band dispersions with Supplementary Eqs. (6) and (9). Then using the DFT calculated values $s_y$ and Supplementary Eqs. (15) and (16) we can find parameters $\alpha$ and $\gamma_1$, using these values and Supplementary Eqs. (10), obtain $\gamma_2$ and $\gamma_3$. The results are as follows: $\delta = -0.13$ eV, $\alpha^+ = 1.91$ eV Å, $\alpha^- = 1.51$ eV Å, $\alpha = -0.18$ eV Å, $\gamma_1 = -1.42$ eV Å, $\gamma_2 = 0.95$ eV Å and $\gamma_3 = 0.09$ eV Å.

## Supplementary Note 3.

### Higher order corrections and deviations from PST

Second-order corrections to the energy $E^{(2)}$ are given by

$$\begin{cases} E_1^{(2)} = E_2^{(2)} = \dfrac{t}{\Delta} k_y^2 \\ E_3^{(2)} = E_4^{(2)} = -\dfrac{t}{\Delta} k_y^2 \end{cases}, \tag{17}$$

where $\Delta = 2\delta$ is the zero-order energy splitting and $t$ is defined as follows

$$\begin{cases} t = |C_1|^2 + |C_2|^2 \\ C_1 = \dfrac{\beta - c_1 \gamma_4 - c_2 \gamma_4 - c_1 c_2 \beta}{\sqrt{(c_1^2+1)(c_2^2+1)}} \\ C_2 = \dfrac{c_2 \beta - c_1 c_2 \gamma_4 + \gamma_4 + c_1 \beta}{\sqrt{(c_1^2+1)(c_2^2+1)}} \end{cases}. \tag{18}$$

Thus, up to the second-order perturbation, the eigenvalues $E_j = E_j^{(0)} + E_j^{(1)} + E_j^{(2)}$ $(j=1-4)$ are given by

$$\begin{cases} E_1 = \delta - \alpha^+ k_x + \dfrac{t}{\Delta} k_y^2 \\ E_2 = \delta + \alpha^+ k_x + \dfrac{t}{\Delta} k_y^2 \\ E_3 = -\delta - \alpha^- k_x - \dfrac{t}{\Delta} k_y^2 \\ E_4 = -\delta + \alpha^- k_x - \dfrac{t}{\Delta} k_y^2 \end{cases}. \tag{19}$$

First-order corrections $\psi_j^{(1)}$ to the eigenstates $\psi_j^{(0)}$ Supplementary (11) produce first-order eigenstates $\psi_j = \psi_j^{(0)} + \psi_j^{(1)}$ $(j=1-4)$ as follows:

$$\begin{cases} \psi_1 = \psi_1^{(0)} + \dfrac{C_1 k_y}{\Delta} \psi_3^{(0)} + \dfrac{C_2 k_y}{\Delta} \psi_4^{(0)} \\ \psi_2 = \psi_2^{(0)} - \dfrac{C_2 k_y}{\Delta} \psi_3^{(0)} + \dfrac{C_1 k_y}{\Delta} \psi_4^{(0)} \\ \psi_3 = \psi_3^{(0)} - \dfrac{C_1 k_y}{\Delta} \psi_1^{(0)} + \dfrac{C_2 k_y}{\Delta} \psi_2^{(0)} \\ \psi_4 = \psi_4^{(0)} - \dfrac{C_2 k_y}{\Delta} \psi_1^{(0)} - \dfrac{C_1 k_y}{\Delta} \psi_2^{(0)} \end{cases}, \tag{20}$$

where $C_1$ and $C_2$ are given by Supplementary Eq. (18).

From Supplementary Eqs. (13) and (20), we see that the perturbed eigenstates are no longer the eigenstates of $\sigma_y$ due to mixing between the doublets. Taking state $\psi_1$, which corresponds to the lowest conduction band in Supplementary Eq. (20), as an example, we find the expectation values of $s_x$ and $s_y$ up to the first order in perturbation theory:

$$\begin{cases} s_x = \frac{1}{2} \frac{\langle \psi_1 | \sigma_x | \psi_1 \rangle}{\langle \psi_1 | \psi_1 \rangle} = q \frac{k_y}{\Delta} \\ s_y = \frac{1}{2} \frac{\langle \psi_1 | \sigma_y | \psi_1 \rangle}{\langle \psi_1 | \psi_1 \rangle} = \frac{1 - c_1^2}{2(c_1^2 + 1)} \end{cases}, \quad (21)$$

where

$$q = \frac{(1 - c_1 c_2) C_1 + (c_1 + c_2) C_2}{\sqrt{(c_1^2 + 1)(c_2^2 + 1)}} . \quad (22)$$

Note that in Supplementary Eq. (21) we omitted the quadratic terms in $k_y$ as they correspond to the higher-order perturbation. We see that within this approximation, the $s_y$ component of the spin remains unchanged and constant, whereas the $s_x$ component becomes nonzero and linear in $k_y$.

Supplementary Eqs. (19) and (21) can be used to obtain the two yet undermined constants in the model Hamiltonian i.e. Supplementary Eq. (2), i.e. $\beta$ and $\gamma_4$. By fitting the DFT calculated band structure along the high-symmetry X-S direction ($k_x = \pi$, $k_y$, $k_z = 0$) we find parameter $t = 0.0559$ eV$^2$ Å$^2$ and by fitting the DFT calculated value of $s_x$ as a function $k_y$ we find parameter $q = -0.139$ eV Å. From $t$ and $q$ using Supplementary Eqs. (18) and (22), we obtain $C_1 = -0.161$ eV Å and $C_2 = 0.173$ eV Å. Substituting $C_1$ and $C_2$ into Supplementary Eq.(18), we finally find $\beta = -0.139$ eV Å and $\gamma_4 = 0.191$ eV Å.

Supplementary Fig. 3 shows results for the spin texture around the CBM calculated from first-principles (Supplementary Fig. 3a) and using our model Hamiltonian within the perturbation approach (Supplementary Fig. 3b). First, it is seen that over a very broad region around the CBM the spin magnitude and orientation is nearly uniform. Second, qualitative comparison between the DFT computed and modeled spin textures reveals excellent agreement. This is remains the case when the comparison is made quantitative. As is evident from Supplementary Fig. 3d, $s_x$ increases linearly with $k_y$ consistent with the model prediction. Supplementary Fig. 3e shows that $s_y$ is nearly independent of $k$, which is again in line with the result of Supplementary Eq. (21). Supplementary Fig. 3c shows the calculated spin-orbit field around the CBM. As expected, the magnitude of the spin-orbit field (arrow length) scales linear with $k_x$ (referenced to the X point) and weakly depends on $k_y$.

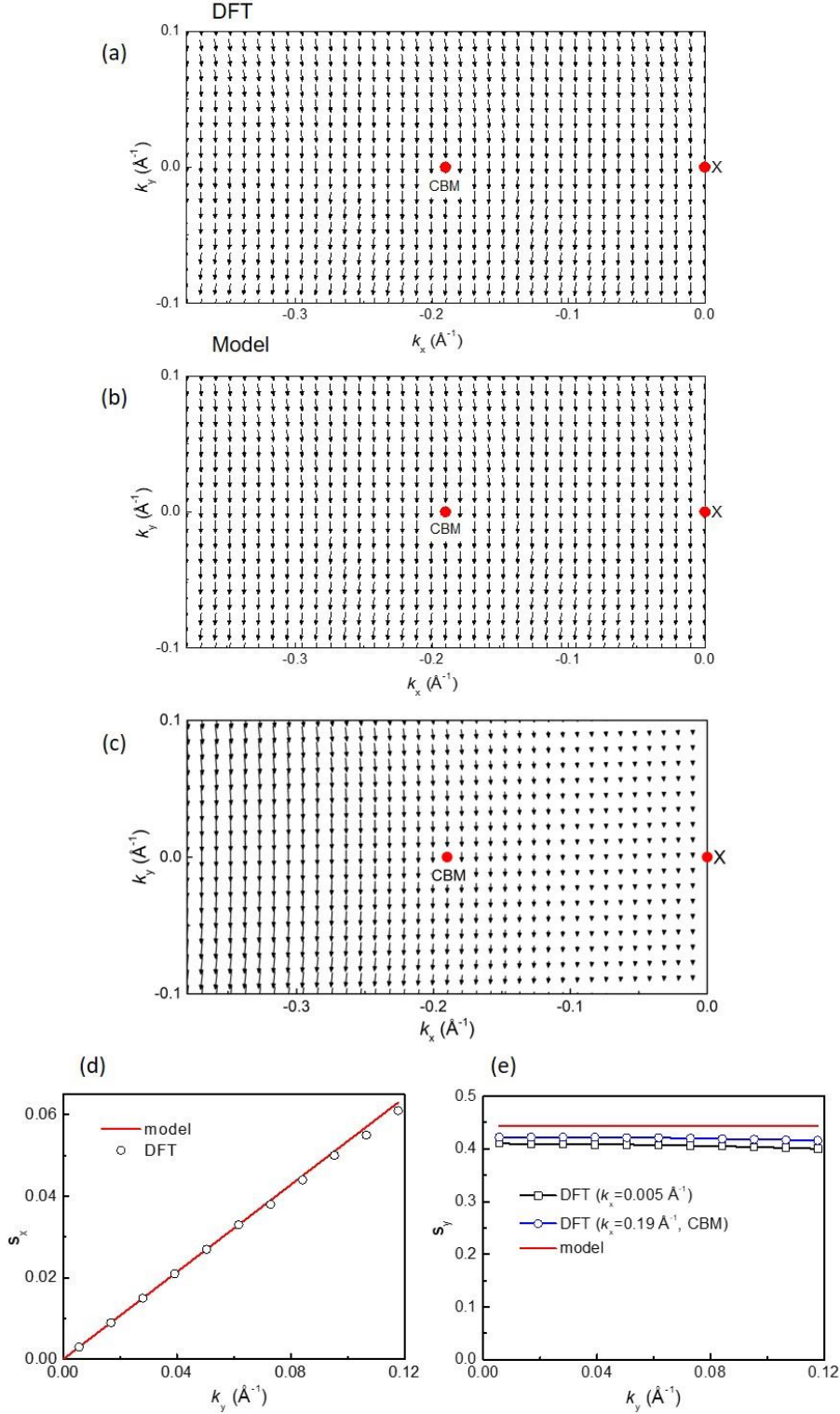

**Supplementary Figure 3. Spin texture around the CBM.** (a, b) Spin textures around the CBM in the $k_z = 0$ plane: DFT results (a) and model results (b) based on Eq. (21). $k_x$ is measured from the X point. (c) Spin-orbit field around the CBM based on the effective Hamiltonian model. (d) Expectation value of the *x* component of the spin, $s_x$. The model results are represented by $s_x = qk_y/\Delta$, where $q/\Delta = 0.535$ Å and $k_x = 0.19$ Å$^{-1}$ corresponding to CBM. (e) Expectation value of the *y* component of the spin, $s_y$. The model results are represented by Supplementary Eq. (21).

## Supplementary Note 4.

### Other PST compounds

Supplementary Fig. 4 shows the crystal structures of BiInS$_3$ and LiTeO$_3$. The orthorhombic $Pna2_1$ structure of BiInS$_3$ (space group No. 33) was proposed in experiment [5] and predicted by first-principles.[6] In our DFT calculations, we used the theoretically predicted lattice constants and atomic coordinates from Supplementary ref.6. The LiTeO$_3$ compound with $Pnn2$ (No. 34) structure was predicted by the Material Project,[7] but the experimental demonstration of this structural phase has not yet been reported. In this work, the lattice constants and atomic coordinates for bulk LiTeO$_3$ were obtained using full structural relaxation with the initial geometry adopted from Supplementary ref.7. Supplementary Table 2 summarizes the lattice constants, atomic positions, calculated polarizations and band gaps for all the three compounds considered in this work.

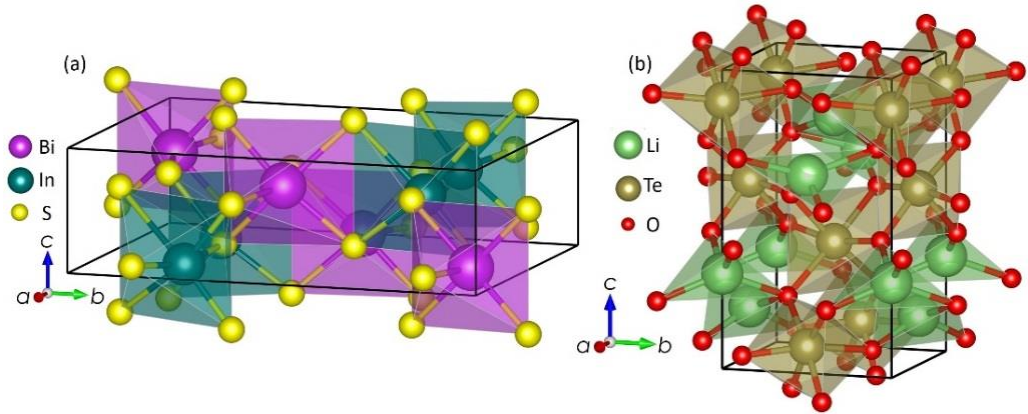

**Supplementary Figure 4**. **Crystal structure of bulk BiInS$_3$ and LiTeO$_3$.** Crystal structure of bulk (a) BiInS$_3$ in the $Pna2_1$ (No. 33) orthorhombic phase and (b) LiTeO$_3$ in the $Pnn2$ (No. 34) orthorhombic phase. Here ($a$, $b$, $c$) axis is concordant with the ($x$, $y$, $z$) axis.

All the three compounds have finite polarization due to broken inversion symmetry. For the orthorhombic crystal system with two perpendicular mirror reflections $M_x$ and $M_y$ (Fig. 1a in the main text), polarization along the $x$ or $y$ direction is forbidden by symmetry. The calculated polarization along the $z$ direction is listed in Table S2. We see that polarizations of BiInO$_3$ and LiTeO$_3$ are similar and comparable to the polarization of conventional ferroelectric oxide BaTiO$_3$, whereas the polarization of BiInS$_3$ is smaller (but still sizeable).

**Supplementary Table 2.** Structural and electronic properties of $BiInO_3$, $BiInS_3$, and $LiTeO_3$ compounds: lattice constants, atomic positions, calculated polarizations $P_z$, band gaps, and energy splittings $\Delta_X$ ($\Delta_Y$) between the two lowest conduction bands at the X (Y) point. The band gaps are calculated using GGA in presence of SOC.

| Compound | Atom | Wyckoff position | x | y | z | $P_z$ (μC/cm$^2$) | Band gap (eV) | $\Delta_X$ (eV) | $\Delta_Y$ (eV) |
|---|---|---|---|---|---|---|---|---|---|
| **$BiInO_3$** | Bi1 | 4a | 0.05955 | 0.00880 | 0.77970 | | | | |
| $a = 5.955$ Å | In1 | 4a | 0.00260 | 0.50110 | 0.00000 | | | | |
| $b = 5.602$ Å | O1 | 4a | 0.05400 | 0.38300 | 0.77300 | 33.6 | 2.26 | 0.26 | 0.09 |
| $c = 8.386$ Å | O2 | 4a | 0.17100 | 0.21800 | 0.44700 | | | | |
| Ref. 8 | O3 | 4a | 0.34400 | 0.63000 | 0.53000 | | | | |
| **$BiInS_3$** | Bi1 | 4a | 0.67016 | 0.44544 | 0.64635 | | | | |
| $a = 10.060$ Å | In1 | 4a | 0.07064 | 0.73238 | 0.59238 | | | | |
| $b = 13.380$ Å | S1 | 4a | 0.39540 | 0.39764 | 0.66734 | 11.4 | 1.13 | 0.03 | 0.05 |
| $c = 3.940$ Å | S2 | 4a | 0.95118 | 0.38956 | 0.59942 | | | | |
| Ref. 6 | S3 | 4a | 0.82959 | 0.80091 | 0.65975 | | | | |
| | Li1 | 2a | 0.00000 | 0.00000 | 0.31812 | | | | |
| | Li2 | 2b | 0.00000 | 0.50000 | 0.66325 | | | | |
| **$LiTeO_3$** | Te1 | 2a | 0.00000 | 0.00000 | 0.89667 | | | | |
| $a = 5.102$ Å | Te2 | 2b | 0.00000 | 0.50000 | 0.08939 | 33.0 | 1.96 | 0.05 | 0.02 |
| $b = 5.293$ Å | O1 | 4c | 0.21217 | 0.65593 | 0.92734 | | | | |
| $c = 8.988$ Å | O2 | 4c | 0.21455 | 0.18574 | 0.06889 | | | | |
| | O3 | 4c | 0.20975 | 0.68100 | 0.23479 | | | | |

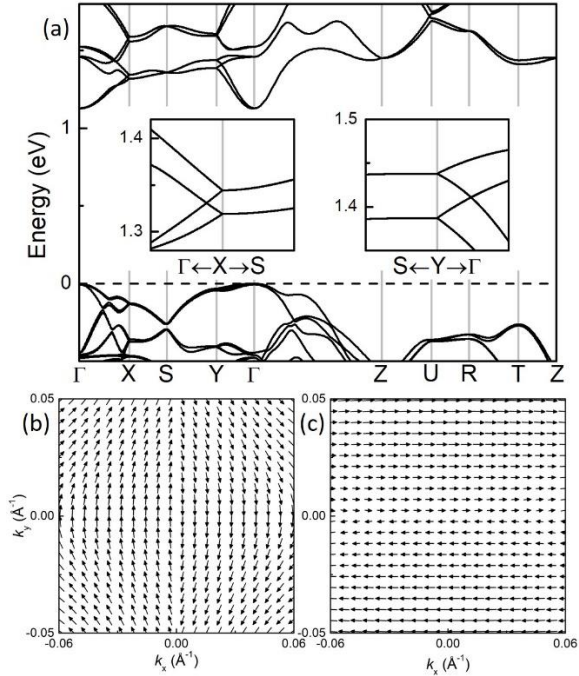
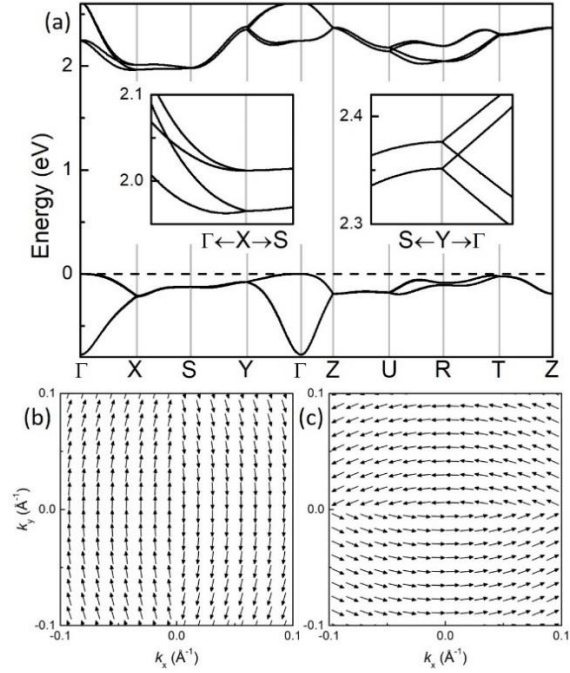

**Supplementary Figure 5. Band structure and spin texture of $BiInS_3$.** (a) Band structure of bulk $BiInS_3$. Inset: zoom-in of band structure around the X and Y points. Spin textures in the $k_z = 0$ plane around the (b) X point and (c) Y point for the lowest conduction band. Note that the wave vector is measured from the X point for (b) and the Y point for (c).

**Supplementary Figure 6. Band structure and spin texture of $LiTeO_3$.** (a) Band structure of bulk $LiTeO_3$. Inset: zoom-in band structure around the X and Y points. (b, c) Spin textures in the $k_z = 0$ plane around the X point (b) and Y point (c) for the lowest conduction band. Note that the wave vector $k_x$ is measured from the X point for (b) and from the Y point for (c). CBM is about 0.01 Å$^{-1}$ from the X point.

Supplementary Figs. 5 and 6 show the band structure and spin texture of BiInS$_3$ and LiTeO$_3$. We can see the similar spin textures around the X and Y points as compared with that for BiInO$_3$. For BiInS$_3$, the CBM is located at the Γ point. The CBM for LiTeO$_3$ is slightly shifted from the X point.

## Supplementary Note 5.

### Symmetry allowed terms up to cubic order in $k$

**Supplementary Table 3.** The symmetry allowed terms in the Hamiltonian within the $k_z = 0$ plane around the X point for crystals of space groups 28, 29, 31-34, 40, 41, 45, and 46 up to cubic order in $k$. The terms are classified according to their effect on PST in zero- and first-order perturbation for the wave function.

| Order in wave vector k | Symmetry allowed terms (all preserve PST in zero order perturbation theory) | Terms which break PST in first order perturbation theory |
|---|---|---|
| Linear | $k_x\sigma_y, k_y\sigma_x, k_x\tau_y, k_x\tau_x\sigma_x, k_x\tau_z\sigma_z, k_y\tau_x\sigma_y$ | $k_y\sigma_x, k_y\tau_x\sigma_y$ |
| Quadratic | $k_x^2, k_y^2, k_xk_y\tau_x, k_x^2\tau_y\sigma_y, k_y^2\tau_y\sigma_y, k_xk_y\tau_y\sigma_x$ | $k_xk_y\tau_x, k_xk_y\tau_y\sigma_x$ |
| Cubic | $k_x^3\sigma_y, k_x^3\tau_y, k_x^3\tau_x\sigma_x, k_x^3\tau_z\sigma_z, k_y^3\sigma_x, k_y^3\tau_x\sigma_y, k_xk_y^2\sigma_y,$ $k_yk_x^2\sigma_x, k_xk_y^2\tau_y, k_xk_y^2\tau_x\sigma_x, k_xk_y^2\tau_z\sigma_z, k_yk_x^2\tau_x\sigma_y$ | $k_y^3\sigma_x, k_y^3\tau_x\sigma_y, k_yk_x^2\sigma_x, k_yk_x^2\tau_x\sigma_y$ |

## Supplementary Note 6.

### Comparison between semiconductor quantum-well structures and non-symmorphic compounds

**Supplementary Table 4.** Comparison of PST and PSH properties in semiconductor quantum-well structures and non-symmorphic compounds.

|  | Semiconductor quantum well | Non-symmorphic compound |
|---|---|---|
| Spin structure | Persistent spin texture | Persistent spin texture |
| Physical origin | Balanced Rashba and linear Dresselhaus parameters | Enforced by non-symmorphic space group symmetry |
| Deviation from PST | Cubic in $k$ Dresselhaus term | Linear in $k$ perturbation term |
| Precession frequency | 0.1–1 THz range | 100 THz range |
| PSH wavelength | μm scale | nm scale |

## Supplementary references